\documentclass[conference]{IEEEtran}
\usepackage[utf8]{inputenc}
\usepackage{newunicodechar, graphicx} 
\usepackage{tikz}
\usepackage{graphicx}
\usepackage{subcaption}
\usepackage{url}
\usetikzlibrary{arrows.meta, positioning, shapes.geometric}

\DeclareRobustCommand{\okina}{%
  \raisebox{\dimexpr\fontcharht\font`A-\height}{%
    \scalebox{0.8}{`}%
  }%
}
\newunicodechar{ʻ}{\okina}

\title{The Community Census and Spatial Visualization Index (CCSVI)}

\author{\IEEEauthorblockN{Aaron McLean}
\IEEEauthorblockA{\textit{University of Hawaii at Manoa} \\
\textit{Laboratory for Advanced Visualization \& Applications}\\
mcleana@hawaii.edu}
\and
\IEEEauthorblockN{Makena Coffman}
\IEEEauthorblockA{\textit{University of Hawaii at Manoa} \\
makenaka@hawaii.edu}
\and
\IEEEauthorblockN{Andy Yu}
\IEEEauthorblockA{\textit{University of Hawaii at Manoa} \\
\textit{ITS - Cyberinfrastructure}\\
andyyu@hawaii.edu}
\and
\IEEEauthorblockN{Scott Nicolas}
\IEEEauthorblockA{\textit{University of Hawaii at Hilo} \\
scottnic@hawaii.edu}
\and
\IEEEauthorblockN{Maja Schjervheim}
\IEEEauthorblockA{\textit{University of Hawaii at Manoa} \\
majaps@hawaii.edu}
\and
\IEEEauthorblockN{Christopher Shuler}
\IEEEauthorblockA{\textit{University of Hawaii} \\
cshuler@hawaii.edu}
\and
\IEEEauthorblockN{Johann Peter Lall}
\IEEEauthorblockA{\textit{University of Hawaii} \\
jpl8484@hawaii.edu}
\and
\IEEEauthorblockN{Sean Cleveland}
\IEEEauthorblockA{\textit{University of Hawaii at Manoa} \\
\textit{ITS Cyberinfrastructure}\\
seanbc@hawaii.edu}
\and
\IEEEauthorblockN{Jason Leigh}
\IEEEauthorblockA{\textit{University of Hawaii at Manoa} \\
\textit{Laboratory for Advanced Visualization \& Applications}\\
leighj@hawaii.edu}
}

\date{April 2026}

\begin{document}

\maketitle
\footnotetext{Project website: \url{https://uh-ci.github.io/ccsvi-dashboard/}}

\begin{figure*}[t]
    \centering
    \includegraphics[width=0.8\textwidth]{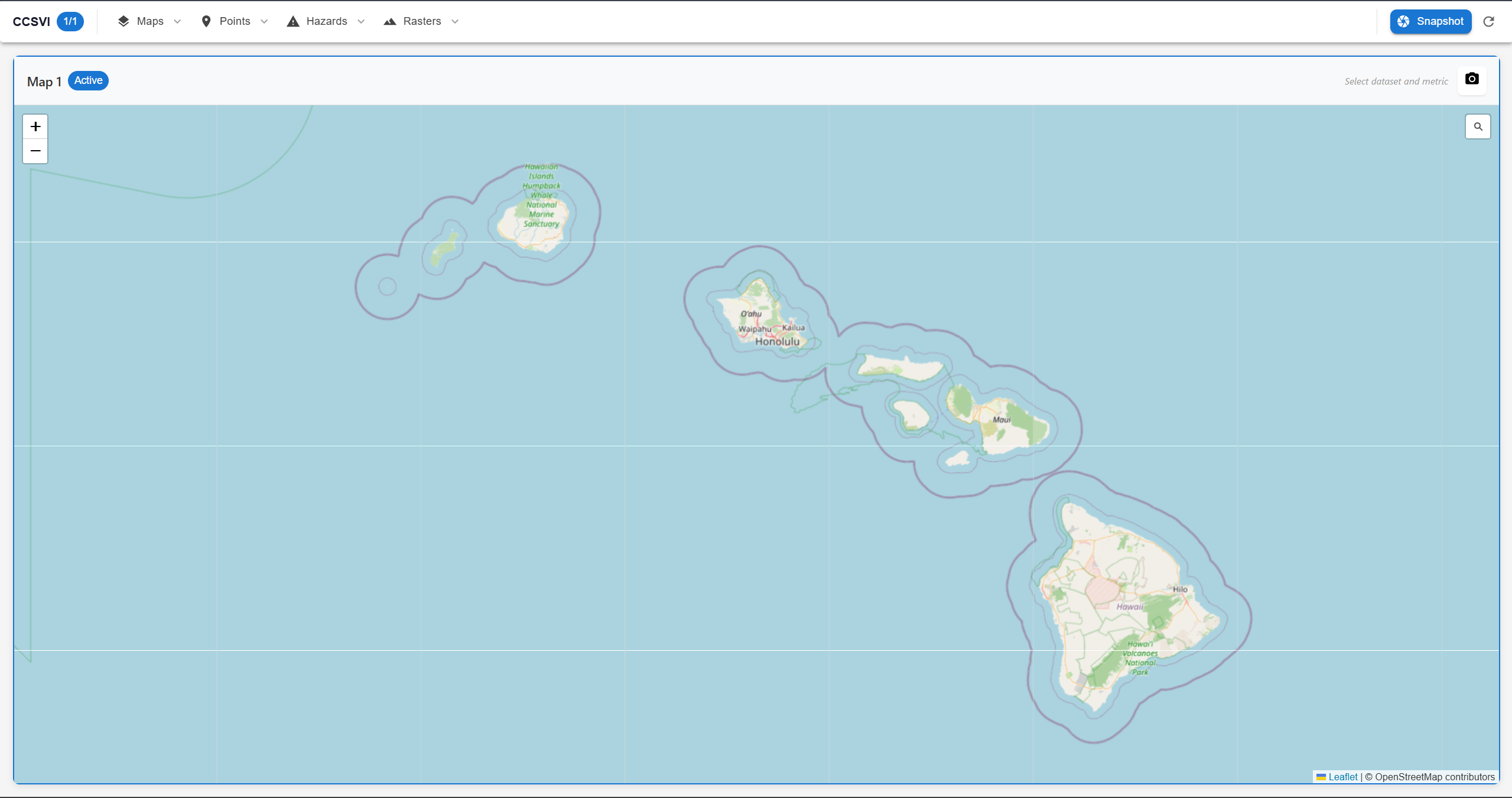}
    \caption{CCSVI Interface. Base interface of CCSVI platform showing the map and control panel with no active data layers. }
    \label{plain}
\end{figure*}

\begin{abstract}
Climate hazards in Hawaiʻi are increasing in both frequency and severity, with varying impacts over vulnerable communities. This paper presents the Community Census and Spatial Visualization Index (CCSVI), a web-based geospatial visualization platform that integrates climate hazard data with socioeconomic and infrastructural datasets. This system enables users to explore the correlation between environmental risks and social vulnerability through interactive mapping and layered data visualizations.

Social vulnerability and climate hazard data are commonly collected individually, this causes the data to be disjointed making it difficult to combine and analyze directly. With data being unrelated when collected, finding direct comparisons and combining the data is difficult resulting in many non-expert users to not understand the data. Additionally, many existing tools focus on only one of these types of data, limiting their interactivity and failing to make any improvements. CCSVI aims to handle the lack of accessible, unified, and interactive systems analyzing the relationship between climate hazards and social vulnerabilities across the state of Hawaiʻi. This support favors assisting decision-makers, researchers, and community members in identifying at-risk populations, improving disaster preparedness, and creating informed climate adaptation strategies.

\end{abstract}

\section{Introduction}
Climate hazards and natural disasters are having an increasing impact on the Hawaiian Islands. Recent severe storms have caused widespread damage, supporting the growing need for both community-level preparedness and coordinated federal support \cite{durepo2026hawaii}. Hawaiʻi consists of diverse communities with varying socioeconomic conditions, geographical features, and infrastructure systems. As a result, the effects of climate hazards are unevenly distributed, resulting in some populations experiencing disproportionately higher levels of risk and vulnerability compared to others. 

However, identifying and quantifying these disparities remains a significant challenge. Climate data are often developed separately from socioeconomic and infrastructural datasets, leading to disconnected information sources. These datasets frequently differ in their spatial resolution, formatting, and update frequencies, making integration both technically complex and tedious. Furthermore, many existing resources and Geographic Information Systems (GIS) present data in a form that is difficult for non-experts to interpret. This causes limitations for decision-makers, community members, and researchers to identify at-risk populations, understand the risks, and develop efficient strategies for disaster preparedness and climate adaptation.  

This project, the Community Census and Spatial Visualization Index (CCSVI), introduces a web-based platform for exploring the relationships between climate hazard exposure, socioeconomic conditions, and infrastructure across the Hawaiian Islands. The system is designed to support data-driven decision-making by helping users better understand how environmental risks intersect with social vulnerabilities, including income distribution, population density, housing conditions, and access to critical infrastructure \cite{coffman2022social}. By integrating these datasets within a single geospatial framework, the platform can identify patterns where environmental hazards and social conditions overlap. These relationships are often difficult to see when each dataset is analyzed separately. This approach enables a more detailed assessment of how vulnerability varies across communities and highlights populations that may experience greater impacts due to the combined effects of environmental exposure and underlying social conditions. 

This system also supports analysis across multiple spatial scales allowing users to compare risk across regions and hazard types through interactive visualizations. This makes it easier to prioritize mitigation efforts, allocate resources more effectively, and support emergency preparedness planning. By presenting complex geospatial and demographic data in an accessible interface, the system reduces barriers to interpretation while preserving analytical depth. This supports more informed context-aware decision-making by planners, policymakers, and community stakeholders. 

The main contributions of this work are to:
\begin{itemize}
    \item Integrate geospatial and socioeconomic datasets into a unified platform
    \item Implementation of optimization techniques for large geospatial datasets
    \item Develop a scalable web-based visualization system for climate vulnerability analysis
    \item Support multi-map comparative analysis with interactive data exploration
\end{itemize}

\section{Background}
This project examined the relationships between social vulnerabilities and climate-related hazards on the Hawaiian Islands. In the field of climate science, there is a gap between scientific analysis and actionable decision-making. Although large volumes of both environmental and socioeconomic data exist, there are few relatable factors for comparing these datasets without a tool to facilitate the analysis. By combining multiple datasets, this study highlighted the disparities between environmental risks and the populations most affected by them.

CCSVI is a future work initiative stemming from the "Social Vulnerabilities with Climate Change in Hawaiʻi" framework developed by Coffman et al. (2022) \cite{coffman2022social}. This report aimed to assess how climate hazards intersected with social vulnerabilities across the Hawaiian Islands by developing a vulnerability index derived from a wide range of socioeconomic and environmental indicators. Key factors identified in the report included income, housing conditions, access to resources, and demographic characteristics. Coffman et al. analyzed how these variables influenced a community's ability to prepare for, respond to, and recover from climate-related events. The report provided a structured methodology for understanding and quantifying vulnerability, as well as guidance for policymakers and community planners.

However, the framework was primarily presented through static analysis, reports, and aggregated visual outputs. While effective for identifying patterns, it limited the ability to dynamically explore relationships between individual variables, compare scenarios, or interact with the data at different spatial levels.

The CCSVI project extended this foundational work by transforming the framework into an interactive visualization system. Led by the same researcher, this project translated the previously theoretical and analytical model into a dynamic, user-driven platform. By integrating geospatial mapping, layered datasets, and interactive tools, CCSVI allowed users to directly explore how social vulnerabilities intersected with specific climate hazards. This transition from a static framework to an interactive system supported more flexible analysis, improved accessibility to complex data, and enhanced the ability to apply vulnerability insights to real-world planning and decision-making contexts.

\begin{figure*}[t]
    \centering
    \includegraphics[width=0.8\textwidth]{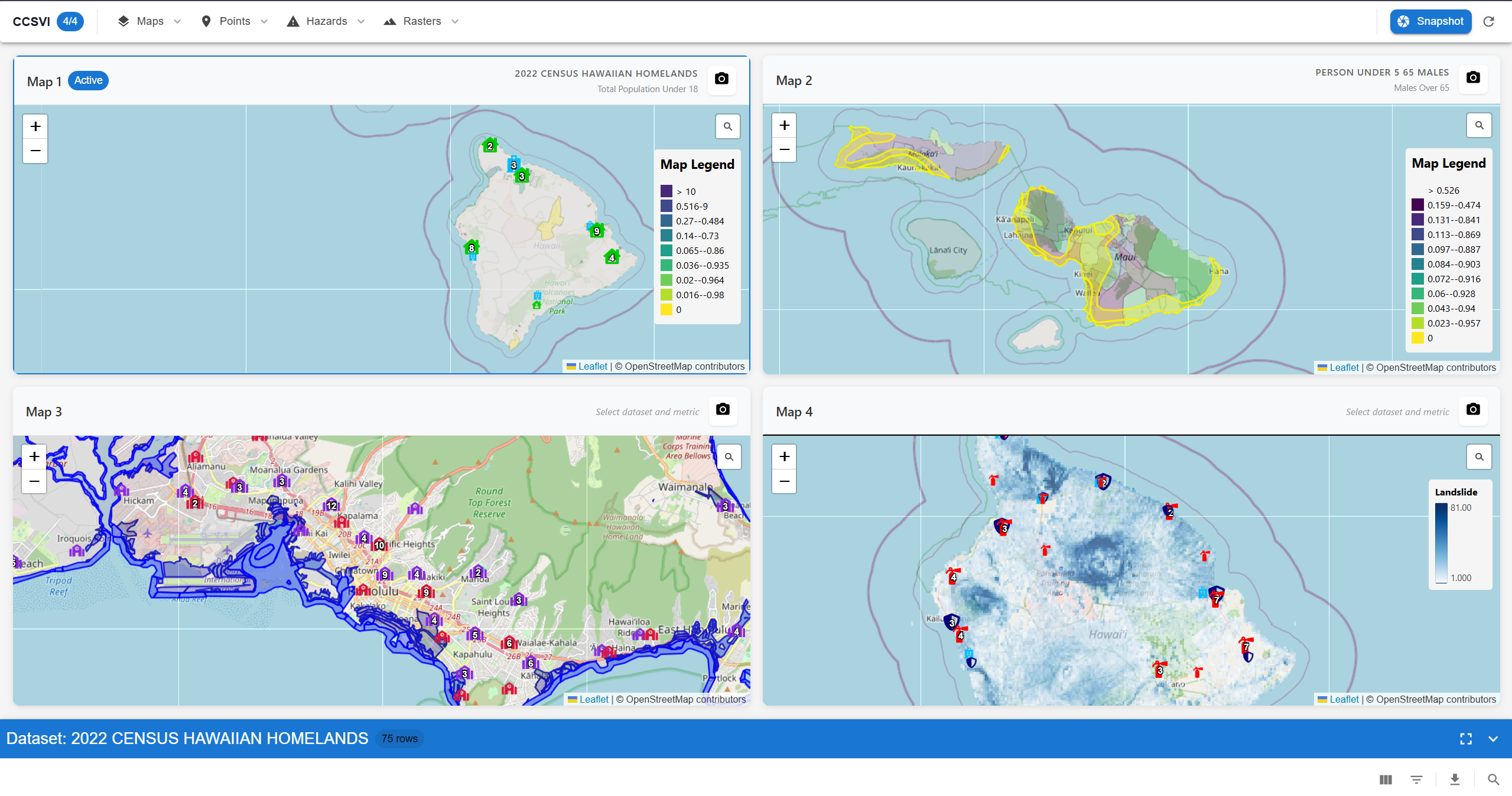}
    \caption{CCSVI Interface. Fully populated CCSVI interface demonstrating multi-map functionality, active data layers, and the table viewer component.}
    \label{FullVisual}
\end{figure*}

\section{Related Work}
\textbf{GIS-based Hazard \& Vulnerability Mapping Systems} There are many GIS resources for the Hawaiian islands, displaying a wide variety of geospatial data and information. Many climate-based projects can be seen displaying geographical data upon a map and providing detailed representations of the chosen dataset. However, while their focus is to present the few climate-based hazards as single representations, this project integrates multiple types of datasets and seek to find additional relationships. 

One example of these GIS resources comes from the "Social Vulnerability Index and Other Related Maps and Data" (SOVI) website\cite{resilientOahuEquityData}. The resources that they present display vulnerability index data for the island of Oʻahu and how they are affected by storms such as Tsunamis, Hurricanes, Flooding, and Sea Level Rise. The CCSVI, in contrast, incorporates all of the Hawaiian islands and had an increasing amount of hazard data. The two projects are similar in ideas, but this project is rapidly updating with fresh data and new information while the SOVI data is getting dated. 

Another example is the Pacific Islands Ocean Observing System (PacIOOS)'s "Sea Level Rise: Hawaiʻi Shoreline Viewer"\cite{pacioosSlrHawaii}. In this interactive mapping tool, the sea level rise vulnerability and adaptation report can be seen for the State of Hawaiʻi. The visualization focuses on the dataset of sea level rise, displaying one layer at a time, and identifying scenarios depending on how much the sea raises. While, the data also includes passive flooding and coastal erosion there are minimal impact assessments on the community. CCSVI combines a similar visual display with additional geospatial data and vulnerability data to present those directly in danger. This viewer only presents the direct impact on locations of the island and not metrics of who or what will be affected by this natural phenomena. 

\section{Data}
There are three main categories of data in the CCSVI project: climate hazards, vulnerability data, and infrastructure data. Climate hazard data is a quantified potential occurrence of physical climate-related events, e.g., flooding hazards, landslides, solar insolation, that cause damage, loss, or disruption. The census vulnerability data are different metrics of community groups within the islands. They measure a community's capacity to prepare for, respond to, and recover from a hazard caused by natural climate hazards. Infrastructure data is important structural buildings, e.g., police stations, hospitals, schools, and bridges. This data focuses on the critical constructs for a community's operation, safety, and economy. By overlapping these datasets to analyze and visualize relationships between the environmental risks, socially vulnerable populations, and different integral structures. 

Climate hazard data is primarily sourced from the Hawaiʻi State Geospatial Data Portal, which is a source that provides GIS datasets for the state. This resource supports improvements in the efficiency and effectiveness of government decision-making through the facilitation of education, coordination, and implementation of GIS mapping technologies \cite{HawaiiGIS}. The majority of the climate hazard and infrastructure datasets were derived from this source; however, additional geospatial data were incorporated from external providers, including NOAA, as well as previously visualized datasets available for the state.

Community vulnerability data were obtained through the United States Census Bureau\cite{censusGov}. These datasets included demographic, economic, and housing-related metrics that were then used to construct the vulnerability indices. The information collected allowed for the identification of populations that were at the greatest risk during climate-related events and received support from targeted planning and resource allocations. 

\section{Data Processing}
CCSVI contains diverse data in various formats, requiring different postprocessing to make them compatible with the visualization framework. The file types processed include Shapefile (SHP), Tagged Image File Format (TIFF/TIF), and CSV. Each format was processed using tools like Quantum Geographic Information System (QGIS) and Python scripts.

\subsection{Shapefile}
Shapefiles (SHP) are vector data storage files for geographical features \cite{templeGuide}. There are three different types of Shapefile including lines, points, and polygons. SHP files are the primary ESRI (Environmental Systems Research Institute) file, which are the systems used in GIS programs like ArcGIS and QGIS, that are what give features their geometry. These files always exist in Shapefiles and are the spatial vector data that form the points, lines, or polygons on a map. Database files (DBF) are the systems that store the structural data of the Shapefile in rows and columns. These hold all of the Shapefile attributes similar to CSV files, but are made for easy operation and connection to GIS software. The last mandatory file is the SHX file, a binary index file that is used to map the relationships between the SHP and DBF files. This file ensures that there is fast searching and loading of the Shapefiles vector data. 

The hazard and points of interest layers of the visualization are primarily Shapefile-based. Due to the varying size of the files, two different methods were taken to process them in order to simplify or reduce the number of points or attributes. 

\subsubsection{QGIS processing}
Most of the hazard layers found as Shapefiles were initially imported into QGIS for basic processing. A copy of the file is created to retain all data for later use in data table visualizations. The first step in looking at the files in the software would be to analyze the attributes if any of the data was irrelevant and could be extracted to minimize the size strain on the processing for visualizations. Once all the attributes were removed, the file would be saved, and further processing would commence. 

The next steps for the Shapefile would be to process it into GeoJSON files, as this is required for loading them into Leaflet maps for the final visualization. Before conversion, the Shapefiles Coordinate Reference System (CRS) must be adjusted to EPSG:4326. EPSG:4326 is a coordinate system used to map geographic coordinates onto a map, and it is the format expected by Leaflet's API. Shapefiles, however, commonly use CRS formats such as EPSG:3857 or EPSG:3750.

In addition to the CRS conversion, the geometry precision of the GeoJSON files needed to be limited to five decimal places, rather than the default precision of 15. Raw datasets used a default of 15 decimal places to support complex spatial calculations, which prevented data loss during these processes. However, both shapefiles and GeoJSON files began to lose accuracy beyond five decimal places, so reducing the precision was necessary to optimize file size and storage space. By limiting the precision, not only can storage be optimized, but the file will also load and process more efficiently. 

After the file is converted, the GeoJSON is loaded into QGIS for geometry simplification. Shapefiles often contain an excessive number of geometry points, which can be reduced without compromising the overall shape or precision of the original data. Simplifying these geometries helps save memory space while maintaining the integrity of the shapes. 

QGIS has a limitation in processing GeoJSON files that exceed a certain size, particularly when the geometry is very dense. As Shapefiles are converted to GeoJSON, some files can grow to several megabytes, making them difficult for QGIS to handle. To address this limitation, a Python script is used to simplify the geometry, ensuring it can be managed efficiently. This QGIS-Python pipeline ensures that files too large for QGIS can still be processed successfully. See the Python method below for more details. 

\subsubsection{Python Map Processing}
The Python-based geospatial processing workflow followed a methodology conceptually similar to that of QGIS, but provided greater transparency and control by explicitly decomposing each transformation step. This approach enabled systematic testing and validation of intermediate outputs, allowing for more informed decisions regarding data optimization. 

The process begins with the ingestion of the shapefile into a geospatial data structure using GeoPandas. Once loaded, the dataset was examined to understand its structural and spatial properties. This included inspecting attribute columns to identify relevant variables, verifying the coordinate reference system (CRS) to ensure spatial consistency, and determining the geometry types (e.g., Polygon or Multi-polygon). Additionally, an initial visualization of the dataset was generated to confirm data integrity and to detect any anomalies in spatial representation. 

Following this exploratory analysis, a geometry simplification procedure was conducted to optimize the dataset for web-based visualization. This step utilized functions derived from Shapely to reduce the number of vertices in each feature while preserving overall spatial correspondence. Rather than applying a single arbitrary tolerance value, multiple tolerance levels were systematically tested as part of the geometry simplification process. In this context, tolerance refers to the maximum allowable deviation between the original geometry and the simplified geometry, effectively controlling how much detail is removed. Lower tolerance values preserve a greater number of vertices and maintain higher spatial accuracy, whereas higher tolerance values result in more aggressive simplification by removing additional points, whereby reducing geometric complexity and file size. 

For each candidate tolerance, the dataset was simplified using functionality provided by Shapely and subsequently exported. This enabled direct comparison of both file size and visual correspondence across different levels of simplification. Through this iterative evaluation, it was possible to observe the trade-off between spatial accuracy and computational efficiency. The selected tolerance value therefore represented an optimal balance, minimizing perceptible visual degradation while significantly reducing the number of vertices and overall file size, making the dataset more suitable for web-based visualization. 

Once an appropriate tolerance value was identified, it was applied to the full dataset as part of the final transformation pipeline. Prior to conversion, the dataset underwent attribute refinement, in which non-essential columns were removed. This step reduced unnecessary data overhead and ensured that only relevant attributes were retained for downstream applications, particularly in web-based environments where payload size directly impacts performance. 

Finally, the processed dataset was converted from its original shapefile format into GeoJSON, a format more suitable for integration with web mapping frameworks such as Leaflet. During this conversion, additional optimizations, such as coordinate precision reduction, were applied to further decrease file size without compromising spatial usability. The resulting GeoJSON file was therefore both structurally streamlined and computationally efficient, making it well-suited for interactive visualization within the broader system architecture.

\subsection{TIFF files}
A Tagged Image File Format or TIFF file, is a raster image format that is maintained by Adobe systems\cite{adobeTIFF}. These files were used to store raster graphics and image data while preserving image quality, though they tended to be larger in size. Raster images were composed of bitmaps that used pixels to store unique colors and tonal information, which together formed an image. Geographical data was applied by assigning a data value to each pixel and then mapping those values to show their distribution across the states. Each pixel in a raster image contained data and information specific to the underlying datasets. The files display climate hazards impact levels across the islands, indicating where impacts would occur and their severity. The particular files used were GeoTIFF files. An important distinction of GeoTIFF from TIFF was that GeoTIFF pixel values represented data measurements, such as elevation. 

The GeoTIFF files were processed using OSGeo4W, which served as a distribution and command-line environment for running geospatial tools. Rather than being an integrated development environment, OSGeo4W provided access to a suite of libraries. and utilities designed to read, analyze, and modify geospatial data formats such as GeoTIFF. While applications like QGIS were capable of working with GeoTIFF files, their large file sizes often made extensive processing slower and more resource-intensive. 

The OSGeo4W environment included numerous geospatial libraries that supported efficient data processing. To improve performance, the files were handled in smaller portions and supplemented with "overviews" when they were not already present. In GeoTIFF files, overviews are precomputed, lower-resolution versions of the original raster data. These allow the system to display simplified versions of the image at smaller scales, reducing the amount of data that must be loaded at once. 

Lowering the resolution at smaller map scales was necessary because full-resolution raster data contains a large number of pixels, many of which are not visually distinguishable when viewed from a distance. By using reduced resolution overviews when zoomed out, the system minimized processing demands and improved rendering speed, while still preserving the ability to display full detail when zoomed in. Larger files typically included multiple overview levels to further optimize performance across different zoom levels. 

This approach significantly improved the loading speed and responsiveness of the raster images by reducing the amount of data processed at any given time, while maintaining the integrity and usability of the underlying TIFF-based data. 

\subsection{CSV}
Comma-Separated Values (CSV) is a simple, widely compatible format for storing and sharing data\cite{adobeCSV}. The vulnerability census data were originally provided as multiple CSV files, each containing structured tabular information. 

Each file first underwent a standard cleaning process, during which special characters and unnecessary spacing were removed to ensure consistency. The data were then restructured to address cases of duplicated or "double" headers. These occurred when multiple columns shared the same base name but were differentiated by suffixes, such as alphabetical or numerical labels (e.g. Population\_A and Population\_B in census datasets). Instead of retaining these as separate columns, the data were normalized by converting them into rows, reducing redundancy and improving consistency across datasets. 

Following this restructuring, the datasets were reviewed to remove unnecessary or conflicting fields. Columns representing estimators and margins of error were excluded because they complicated the process of joining multiple datasets. These values often differ in structure, scale, or interpretation across sources, making direct comparisons unreliable and integration more complex. Additionally, such statistical fields were less suitable for visualization, where clear, comparable metrics were prioritized. Each dataset then underwent a final review to eliminate any remaining superfluous information. 

After processing, all CSV files were consolidated into a single JSON file to simplify data handling. JSON was selected because it provided a flexible, hierarchical structure that better supported nested and grouped data compared to most CSV tables. It was also more efficient for visualization workflows, as it allowed all processed data to be loaded at once rather than requiring multiple file reads. A traditional database was not used at this stage because the project remained in development, and rapid iteration was prioritized. In this context, the JSON file effectively served as a temporary data store, enabling quick access and modification while maintaining a structured representation of the processed TIFF-related datasets. 

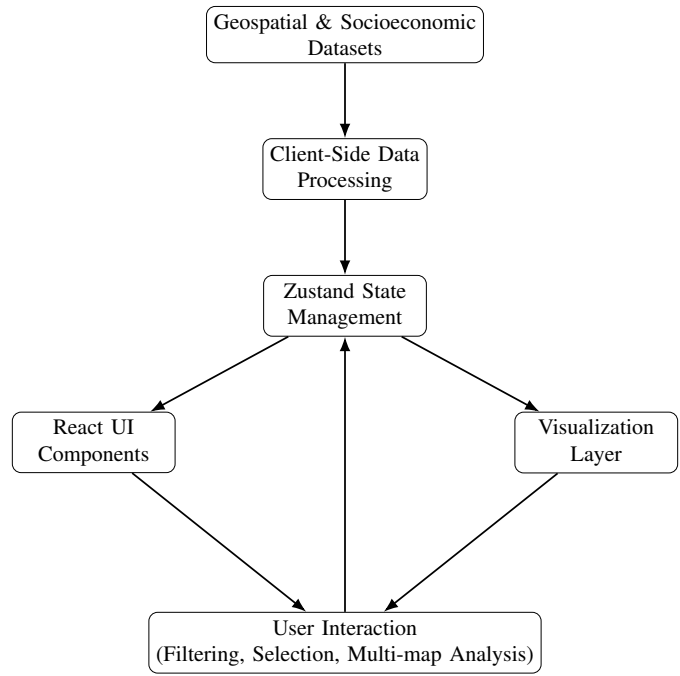
\begin{figure*}[t]
\centering
\resizebox{\columnwidth}{!}{%
\begin{tikzpicture}[
    node distance=1.2cm and 1.4cm,
    box/.style={rectangle, draw, rounded corners, align=center, minimum width=2.6cm, minimum height=0.9cm},
    arrow/.style={-{Latex}, thick}
]

\node[box] (data) {Geospatial \& Socioeconomic\\Datasets};

\node[box, below=of data] (processing) {Client-Side Data\\Processing};

\node[box, below=of processing] (state) {Zustand State\\Management};

\node[box, below left=of state] (ui) {React UI\\Components};

\node[box, below right=of state] (viz) {Visualization\\Layer};

\node[box, below=of state, yshift=-3.2cm] (user) {User Interaction\\(Filtering, Selection, Multi-map Analysis)};

\draw[arrow] (data) -- (processing);
\draw[arrow] (processing) -- (state);

\draw[arrow] (state) -- (ui);
\draw[arrow] (state) -- (viz);

\draw[arrow] (ui) -- (user);
\draw[arrow] (viz) -- (user);

\draw[arrow] (user.north) -- (state.south);

\end{tikzpicture}%
}
\caption{High-level architecture of the CCSVI platform showing data flow through processing and Zustand state management into React UI and visualization components, with direct user-driven state updates.}
\label{fig:ccsvi_architecture}
\end{figure*}

A simplified overview of the system architecture is shown in Fig. ~\ref{fig:ccsvi_architecture}.

\section{CCSVI Architecture}
The CCSVI platform is currently implemented as a modern web application built using React, an industry-standard component-based web framework. The architecture follows a modular, component-driven structure that separates user interface components, application state management, and data visualization logic. 

At a high level, the system is composed of three primary layers: the user interface layer, the state management layer, and the data visualization layer. The user interface layer is implemented in React and is responsible for rendering the interface and handing any user interactions. Zustand is used as a lightweight global state management solution to maintain application state across components, enabling efficient updates without excessive re-rendering. 

The data flow within the system begins with geospatial and socioeconomic datasets. First, the data was processed, geospatial to GeoJSON/GeoTIFF, and socioeconomic to JSON files. JSON files are used in place of a database for an efficient loading method once a dataset has been processed. Once processing has completed and the JSON files are stored, and handled by Zustand state stores. These datasets are then consumed by the visualization components, which render the interactive maps and layered visual outputs. User interactions, such as filtering or selecting data layers, trigger state updates that propagate through the system and dynamically update the visualization in real time. 

This architecture was chosen to prioritize responsiveness and scalability when handling several large geospatial datasets simultaneously. By maintaining computation and rendering on the client side and leveraging a centralized state store, the system reduces unnecessary data duplication and ensures consistent synchronization between visual components.

\section{CCSVI Design}
The web platform was designed to display multiple metrics simultaneously by integrating several datasets and maps. Each dataset was independently toggleable and organized according to its respective elements. As shown in Fig. ~\ref{FullVisual}, all components of the visualization were combined to enable concurrent display and comparison. The census data is visualized per dataset and metric. Hazards are stackable and displayed with different degrees of severity (the only exception is hazards displayed as rasters, as only one of these layers is displayable at one time). Points of interest allow for multiple to be chosen at one time. There are three main features of the CCSVI visualization: Maps, Control panel, and Table viewer. The visualization is still in progress, and the design is having consistent updates.

\begin{figure*}[t]
    \centering
    \includegraphics[width=0.9\textwidth]{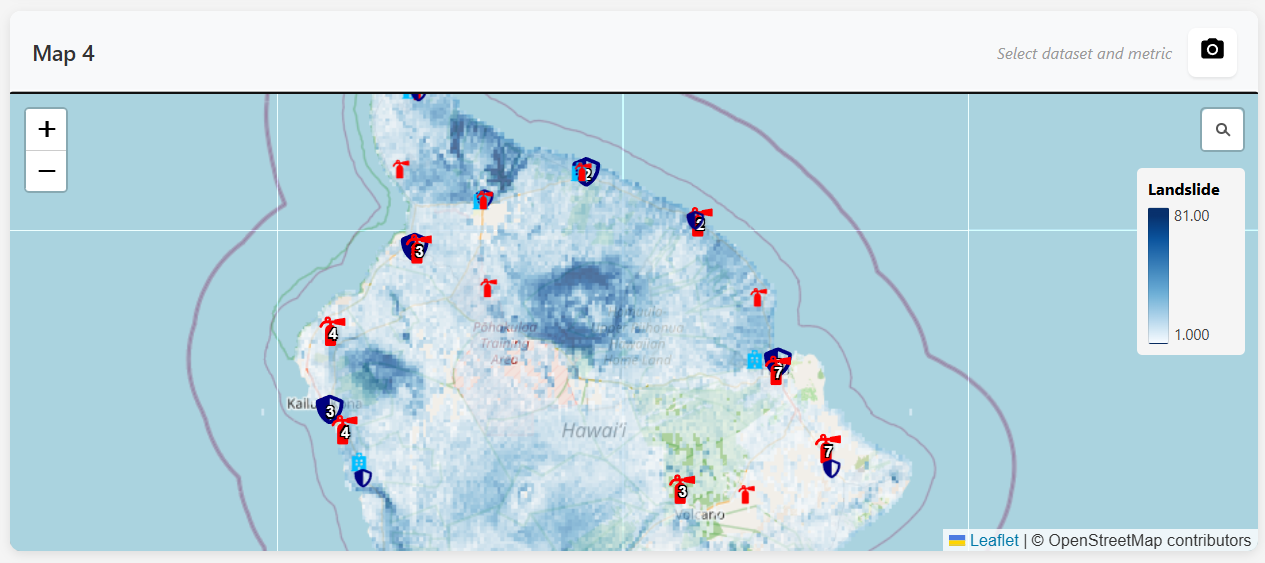}
    \caption{This map illustrates the landslide-prone areas and proximity to emergency response infrastructure on Hawaiʻi Island, examining the potential response challenges}
    \label{map4}
\end{figure*}
\begin{figure*}[t]
    \centering
    \includegraphics[width=0.7\textwidth]{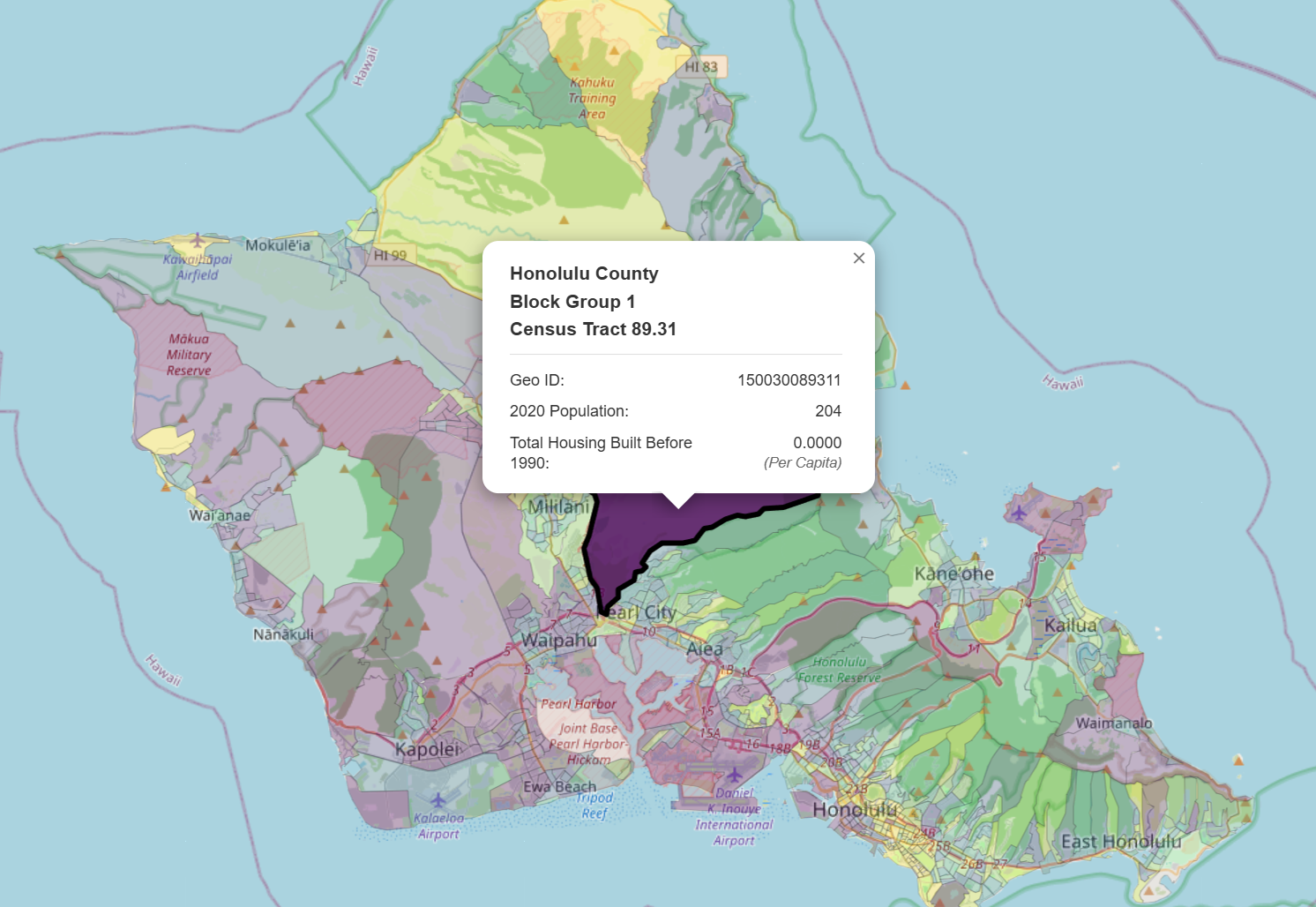}
    \caption{This map illustrates the block grouping for the age of structure dataset}
    \label{blockgroup}
\end{figure*}

\subsection{Map}
The map constituted the primary component of the visualization, displaying the various metrics. The implementation utilized Leaflet as the mapping library for rendering and interaction. To support the simultaneous analysis of multiple metric scenarios, a multi-map function was developed, allowing users to open and control up to four maps concurrently. 

Each map was designed to display a single vulnerability indicator, along with multiple points of interest, multiple climate hazards derived from shapefiles, and one climate hazard represented as a raster layer. This configuration enabled users and agencies to compare different hazard scenarios while also identifying the populations most likely to be affected, thereby supporting more informed planning and decision-making.

The census data were displayed at the block group level using shapefiles with each dataset and corresponding metric maintained as separate layers. These layers were rendered on the map as polygon features, allowing users to visually distinguish geographic boundaries associated with each data value. When a user selected a specific block group, the map automatically zoomed to that feature and generated a pop-up containing relevant attribute data drawn from the associated dataset, enabling more detailed inspection at a localized level. 

Color schemes were applied to represent variations in metric values across regions. These schemes were carefully selected to ensure clear visual differentiation between value ranges while remaining accessible to a broad range of users, including those with color vision deficiencies. Multiple color options were provided to improve readability under different viewing conditions and user preferences, supporting more effective interpretation of the data. 

\begin{figure*}[!t]
    \centering

    \begin{subfigure}{0.9\textwidth}
        \centering
        \includegraphics[width=\textwidth]{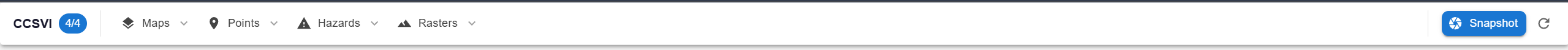}
        \caption{Control panel interface}
        \label{fig:CPInterface}
    \end{subfigure}

    \vspace{0.4cm}

    \begin{subfigure}{0.22\textwidth}
        \centering
        \includegraphics[width=\textwidth]{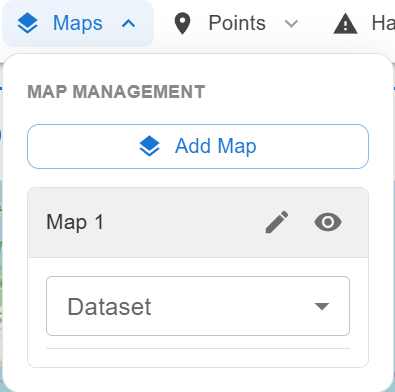}
        \caption{Map selection menu}
        \label{fig:map_menu}
    \end{subfigure}
    \hfill
    \begin{subfigure}{0.22\textwidth}
        \centering
        \includegraphics[width=\textwidth]{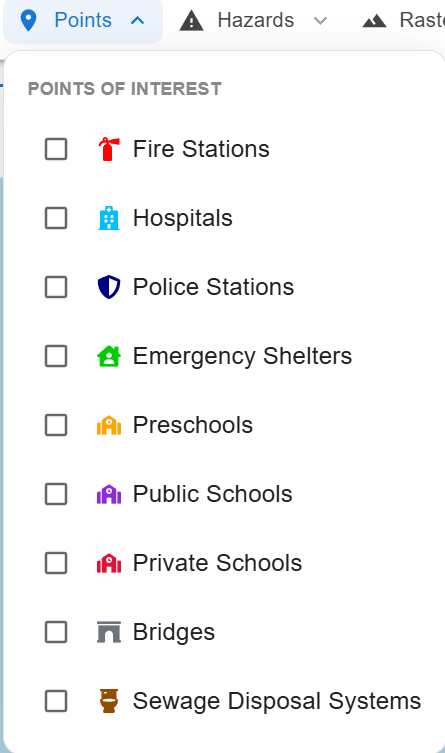}
        \caption{Points of interest menu}
        \label{fig:poi_menu}
    \end{subfigure}
    \hfill
    \begin{subfigure}{0.22\textwidth}
        \centering
        \includegraphics[width=\textwidth]{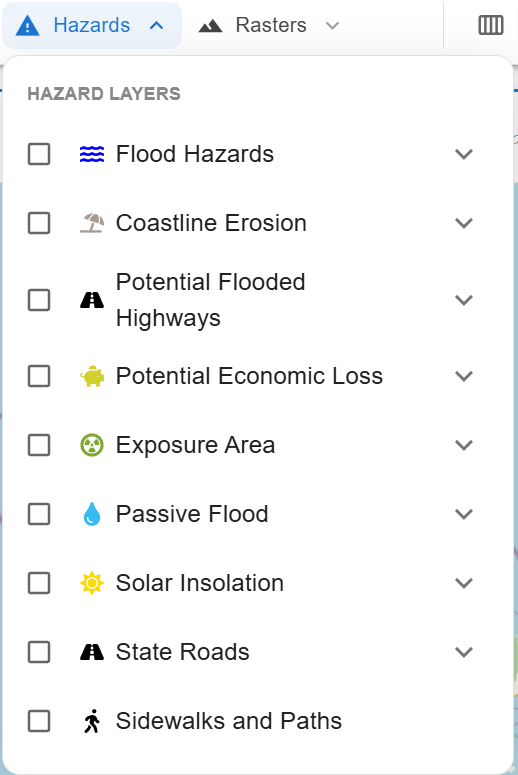}
        \caption{Hazard (shapefile) menu}
        \label{fig:shapefile_menu}
    \end{subfigure}
    \hfill
    \begin{subfigure}{0.22\textwidth}
        \centering
        \includegraphics[width=\textwidth]{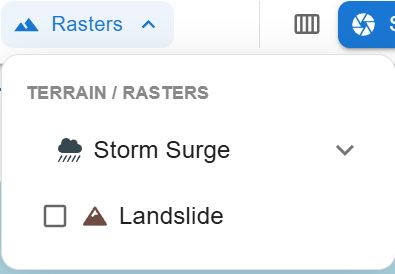}
        \caption{Hazard (raster) menu}
        \label{fig:raster_menu}
    \end{subfigure}

    \caption{Control panel and associated drop-down menus. The interface (top) provides access to four data categories: map layer, points of interest, hazard shapefiles, and hazard rasters (bottom). Each drop-down presents the available layers for its respective data type, while snapshot and refresh controls are located on the right}
    \label{fig:controlPanel}
\end{figure*}

The selected dataset and metric were displayed in the top-right corner of the map's title bar, allowing users to quickly identify the active layer and maintain context while interacting with the visualization. 

Points of interest were represented as spatial point features on the map, corresponding to infrastructure-related locations such as schools, police stations, fire stations, and bridges. Each category was displayed using a distinct icon directly associated with the type of infrastructure, allowing users to quickly identify and differentiate between feature types.

To improve performance and maintain visual clarity, a clustering system was implemented for point features at lower zoom levels. When the map was zoomed out, individual points were aggregated into clusters and displayed as a single marker labeled with the number of contained features. This approach was necessary because displaying all individual points simultaneously at wider extents resulted in excessive visual clutter and reduced the readability of underlying map layers. As users zoomed in, these clusters dynamically separated back into their individual point features, restoring full detail and enabling precise spatial identification.

Data displayed on the maps were accompanied by interactive pop-ups associated with each block group feature. In this context, a block group referred to a geographic unit used in census data, and in Leaflet it was implemented as a polygon layer containing structured attribute information. When a user selected a block group, a pop-up was generated displaying the relevant metric values for that geographic unit, along with associated hazard or infrastructure data where applicable. These pop-ups provided a direct method for interpreting spatial data without requiring users to navigate through external tables or layer panels, as all key attribute information was accessible directly from the map interface.

\begin{figure*}[!t]
    \centering
    \begin{subfigure}{0.45\textwidth}
        \centering
        \includegraphics[width=\textwidth]{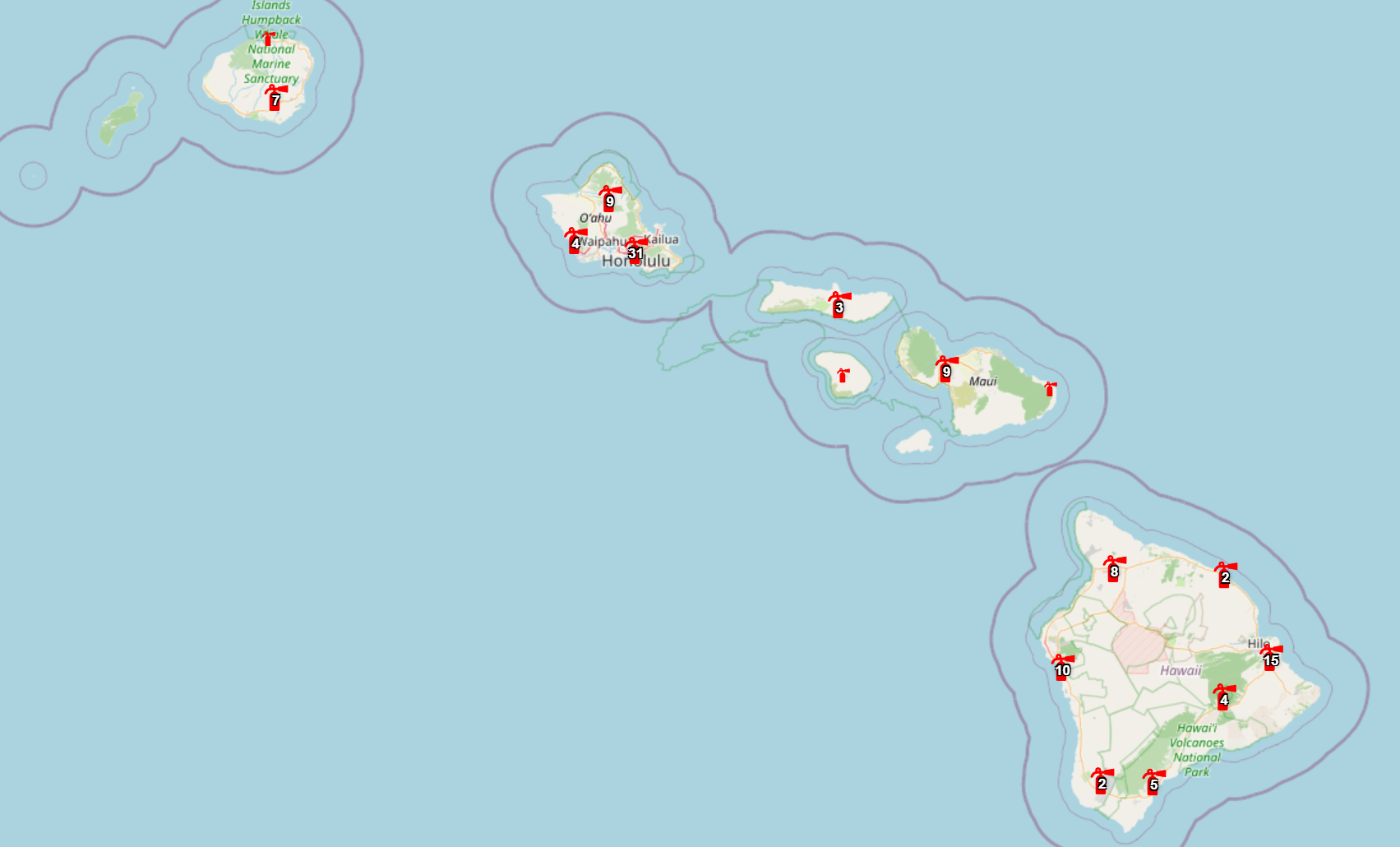}
        \caption{Clustering of fire stations across the state of Hawaiʻi} 
        \label{fig:cluster}
    \end{subfigure}
    \hfill 
    \begin{subfigure}{0.45\textwidth}
        \centering
        \includegraphics[width=\textwidth]{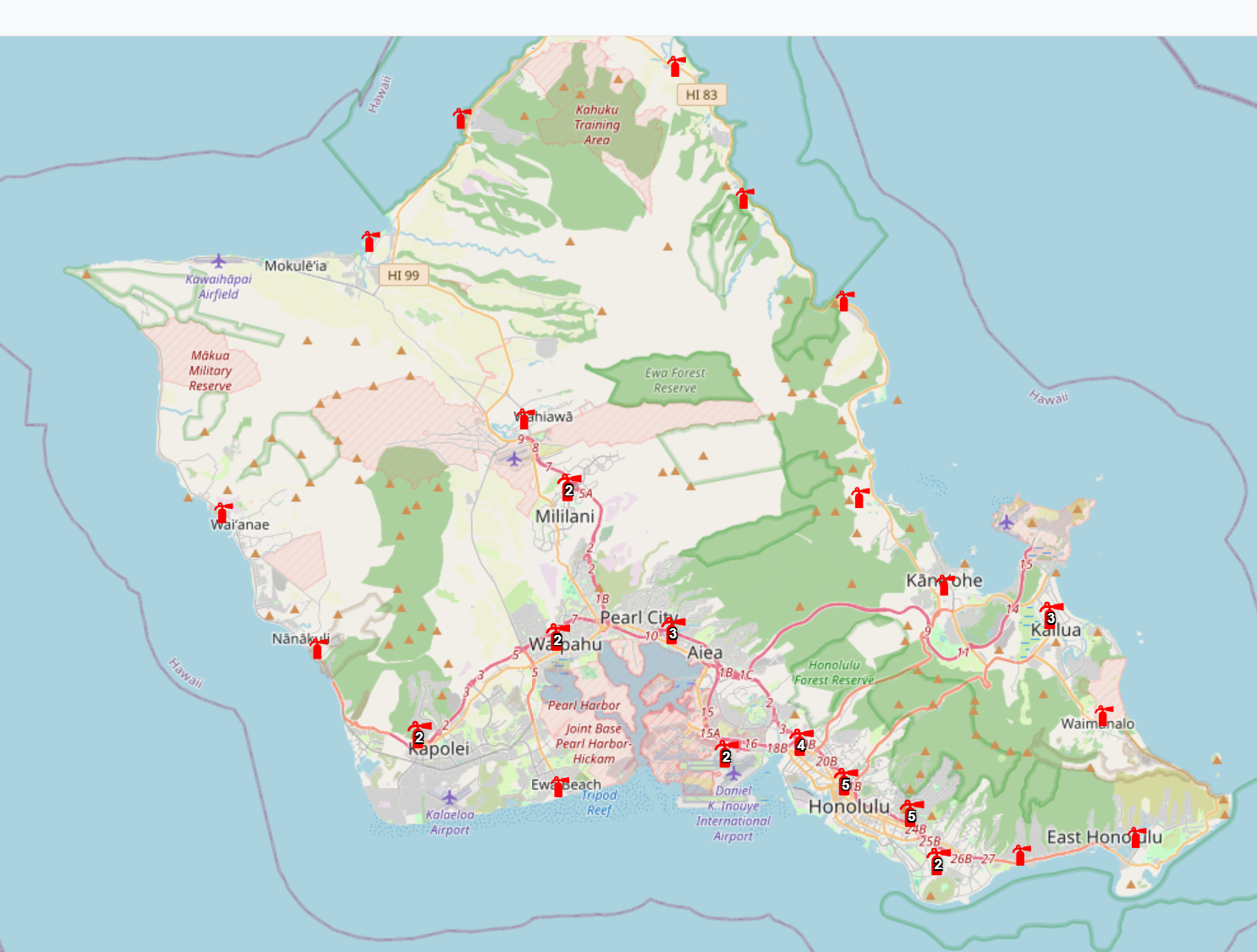}
        \caption{Zoomed-in clustering of fire stations on the island of Oʻahu} 
        \label{fig:ZoomCluster}
    \end{subfigure}
    \caption{Clustering of fire stations enhances map readability by reducing visual clutter.}
    \label{fig:totalCluster}
\end{figure*}

An additional feature of the application was the snapshot function, which allowed users to capture the current state of a map view as a static image. This snapshot included all active layers, applied metrics, and the current zoom level and spatial extent. By preserving the exact visual configuration of the map at a given moment, this feature supported documentation, comparison, and use in presentations or reports. Each map instance contained its own snapshot control, enabling independent capture in a multi-map environment and supporting side-by-side analysis of different scenarios. 

A search function was also implemented to improve spatial navigation. Located in the upper-right corner of the map interface, the search tool allowed users to input a specific address or location. Upon selection, the map automatically zoomed to the corresponding area and placed a marker at the identified location, enabling rapid geographic referencing within the dataset.

\subsection{Control panel}

In order to interact with the map and its data, a control panel was used to toggle individual data layer sets on and off. The control panel was implemented as a bar along the top of the map section. Each set of layers was organized into tabs: Maps, Points, Hazards, and Rasters. When a tab was selected, the corresponding data section was expanded, allowing users to activate or deactivate the required layers.

The control panel design was structured to separate different data types and file formats into distinct categories, ensuring that vector-based data (such as point and shapefile-derived hazard layers) and raster-based data (such as GeoTIFF layers) were managed independently. This separation improved clarity in layer management and reduced the likelihood of user confusion when working with multiple dataset types simultaneously. In addition, the control panel was positioned as a top navigation bar rather than a side panel to preserve horizontal map space and minimize the screen area occupied by interface elements, thereby maximizing the usable viewing area for spatial analysis.

The Maps tab allowed users to create new map instances (up to a maximum of four) to visualize and compare individual census metrics. Within this tab, users could also hide or delete map instances, enabling flexible comparison across different analytical scenarios. The Points tab contained all point-of-interest layers representing infrastructure data. When activated, these points were dynamically clustered based on map zoom level to reduce visual clutter and improve performance at wider extents. The Hazards and Raster tabs contained climate-related layers, with hazards derived from shapefile data and rasters derived from GeoTIFF files.

The control panel also included a refresh function that reset the application to its initial state. This feature was included to allow users to quickly clear all active layers, map instances, and configurations, restoring a consistent baseline for further analysis. This was particularly important in an exploratory workflow, where users frequently tested multiple layer combinations and required a fast method to restart analysis without manually undoing changes.

\begin{figure*}[t]
    \centering
    \includegraphics[width=0.8\textwidth]{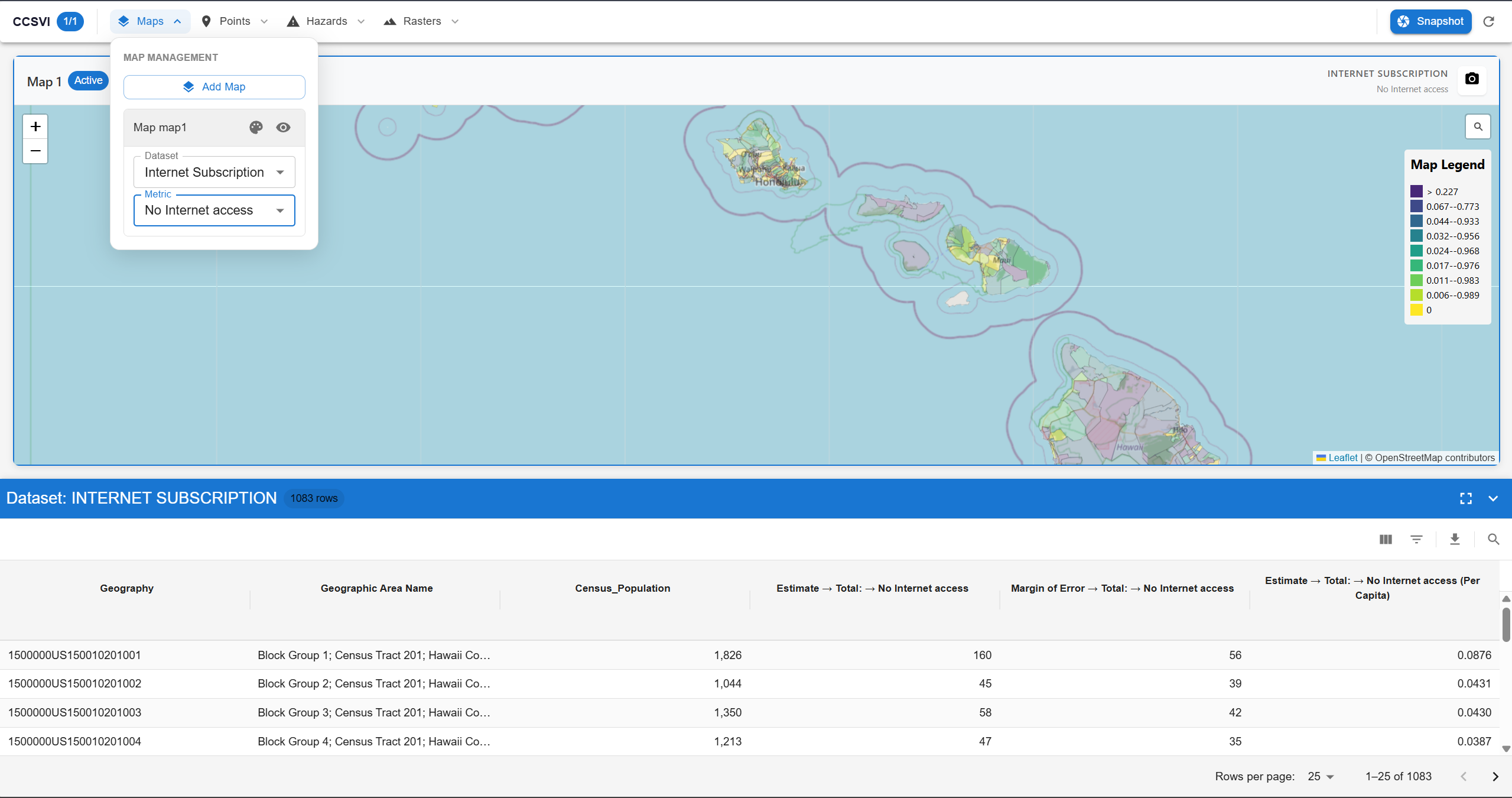}
    \caption{ Table viewer displaying dataset values alongside the geospatial visualization. }
    \label{TableView}
\end{figure*}
\begin{figure*}[t]
    \centering
    \includegraphics[width=0.8\textwidth]{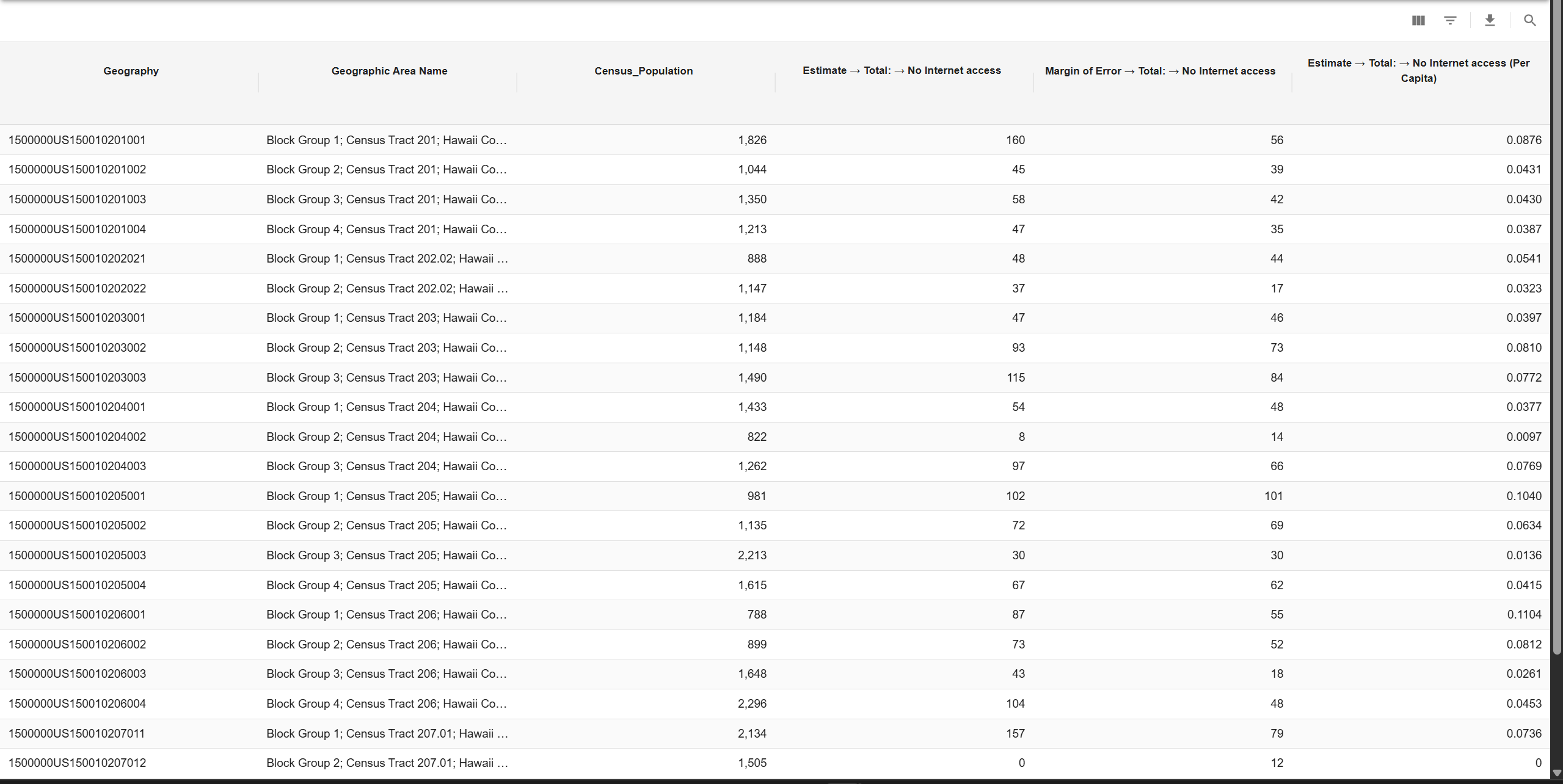}
    \caption{ Expanded table viewer showing detailed dataset records for analysis. }
    \label{fullTableView}
\end{figure*}

\subsection{Table viewer}
This project did not focus exclusively on map-based visualization but also incorporated a complementary table viewer direct interaction with the underlying dataset. The table viewer displayed the metric data associated with the selected dataset and map configuration. After a dataset was selected through the control panel, the table viewer was generated, as shown in FIG. ~\ref{TableView}.

The primary function of the table viewer was data interrogation and, most importantly, data export. Users were able to sort, search, select, and export records. Sorting was enabled by clicking column headers, allowing values to be ordered in ascending or descending order. The search function allowed users to isolate specific values and identify vulnerable communities based on selected metrics. Selecting a row highlighted the corresponding feature on the map, automatically zooming to its location and displaying its associated attribute information for cross-referencing between tabular and spatial views. 

A key feature of the system was the ability to export datasets directly from the table viewer. Exported data reflected the current state of the table, meaning any applied filters or searches were preserved in the output. This allowed users to generate refined datasets tailored to specific analytical needs rather than exporting only the full raw dataset. Each vulnerability dataset could therefore be exported either in full or as a filtered subset, supporting downstream analysis and reporting workflows. 

In addition to the standard view, the table viewer could be expanded to display a larger portion of the dataset simultaneously, as shown in Fig. ~\ref{fullTableView}. This extended view was intended to support more detailed inspection of rows and attributes when visual comparison was less important than direct data analysis. 

\subsection{URL Sharing}
This project has the capability of URL Sharing. URL Sharing is the ability for the website to have multiple user interactions done to the display, adding more maps, layers, and other interactions, then taking the URL and when opened on another computer has all of the same layers and interactions active. The use of this component is for different users, such as allowing other researchers to see your view of the interactions of the data, or displaying an interesting metric combination to another user on another island. The ability to share your findings to other collaborators with a link gives a better idea of what is being visualized compared to sharing a single snapshot of the data that may leave them unsure of what they are looking at. 

\section{Performance Evaluation}
Due to the large size of some datasets, several optimization techniques were applied to improve performance and ensure the system remained responsive and scalable. Geospatial shapefiles were simplified by reducing coordinate precision from 15 to five decimal places. At higher precision levels, particularly beyond five decimal places, there was no meaningful improvement in spatial accuracy for visualization purposes, while storage requirements and rendering complexity increased significantly. Reducing precision therefore eliminated unnecessary data overhead without compromising visual integrity. 

In addition, geometric simplification was applied using a tolerance-based method to reduce inner geometry complexity. Inner geometry points refer to redundant vertices located along straight or minimally curved segments of a shape that do not materially affect its overall structure. These points were removed where possible to streamline the dataset while preserving the original form of each feature. The degree of simplification varied across datasets depending on their initial complexity, with reductions ranging from approximately 40\% to over 90\%, as shown in Table ~\ref{table1}. This reduction significantly improved rendering efficiency while maintaining accurate spatial representation of all features. 

Raster datasets were optimized through the use of GeoTIFF overviews. Overviews are pre-generated lower-resolution representations of original raster data that are displayed as reduced zoom levels. This approach reduced the number of pixels that needed to be rendered at once, ensuring smoother performance when navigating the map. Without overviews, full-resolution rasters at low zoom levels could take several minutes to render or, in some cases, cause system instability due to memory overload. By incorporating overviews, rendering became immediate and stable across all zoom levels, substantially improving both performance and user experience. 

Together, these optimizations significantly improved the responsiveness of the visualization system and enabled smoother interaction across multiple active datasets, even under high computational load.

\begin{table}[t]
\caption{Geometry Removed}
\begin{center}
\begin{tabular}{|c|c|c|c|}
\hline
\textbf{Layer name} & \textbf{unfiltered} & \textbf{filtered} & \textbf{percent removed} \\
\hline
     slr\_Potent\_fld\_hws\_3pt2ft & 3,122    & 1,477   & 52.69058296 \\
     \hline
    slr\_Potent\_fld\_hws\_2pt0ft & 1,554    & 726     & 53.28185328 \\
    \hline
    slr\_Potent\_fld\_hws\_1pt1ft & 986      & 495     & 49.79716024 \\
    \hline
    slr\_Potent\_fld\_hws\_0pt5ft & 652      & 386     & 40.79754601 \\
    \hline
    slr\_passive\_fld\_3pt2ft     & 2,948,014 & 259,675 & 91.19152758 \\
    \hline
    slr\_passive\_fld\_2pt0ft     & 2,175,961 & 150,378 & 93.08912246 \\
    \hline
    slr\_passive\_fld\_1pt1ft     & 1,825,105 & 130,609 & 92.84375419 \\
    \hline
    slr\_passive\_fld\_0pt5ft     & 1,694,281 & 137,456 & 91.88706006 \\
    \hline
    slr\_exposure\_area\_3pt2ft   & 2,035,478 & 164,078 & 92.88321445 \\
    \hline
    slr\_exposure\_area\_2pt0ft   & 1,768,409 & 127,812 & 92.77284636 \\
    \hline
    slr\_exposure\_area\_1pt1ft   & 1,512,318 & 101,005 & 93.32117798 \\
    \hline
    slr\_exposure\_area\_0pt5ft   & 1,378,619 & 90,820  & 93.41224482 \\
    \hline
    slr\_cstl\_erosn\_3pt2ft      & 23,281   & 1,794   & 92.44914484 \\
    \hline
    slr\_cstl\_erosn\_2pt0ft      & 22,978   & 1,900   & 91.73122172 \\
    \hline
    slr\_cstl\_erosn\_1pt1ft      & 22,932   & 2,143   & 90.65497937 \\
    \hline
    slr\_cstl\_erosn\_0pt5ft      & 23,533   & 2,512   & 89.32562784 \\
    \hline
\end{tabular}

\label{table1}
\end{center}
\end{table}

\section{Future Work}
\subsection{HCDP \& Other Layers}
Future development aimed to broaden both climate and socioeconomic coverage to improve the robustness of vulnerability analysis. This included the addition of more high-resolution hazard layers, such as precipitation extremes, heat exposure, wildfire risk, and sea-level rise projections, alongside refined coastal and inland flooding models. These additions were intended to better capture the full range of environmental risks affecting the Hawaiian Islands and to support more detailed spatial and temporal comparisons. 

On the socioeconomic side, further census-derived and administrative datasets were planned for integration to improve representation of population sensitivity and adaptive capacity. These expansions were intended to strengthen the analytical depth of the system by incorporating additional indicators of social vulnerability and infrastructure dependence, enabling more comprehensive multi-layered risk assessments. 

A key planned enhancement was the integration of the Hawaiʻi Climate Data Portal (HCDP), a centralized platform providing access to high-quality, standardized climate datasets\cite{hawaiiCliamtePortal}.  HCDP supported improved consistency and reliability in climate information by aggregating validated observational and modeled datasets. Within CCSVI, HCDP data would be used to strengthen predictive capabilities by supplying high-resolution climate variables such as rainfall, temperature, humidity, and related atmospheric indicators. These variables were particularly valuable for identifying patterns and conditions associated with hazard formation and intensification. By incorporating HCDP data, the system would be better equipped to support predictive analysis of climate-driven events and improve the temporal understanding of risk across the islands. 

Overall, the goal of these future enhancements was to transition the system from a foundational dataset collection into a more complete, predictive vulnerability modeling framework capable of supporting long term planning and decision-making.

\subsection{Database and AI Queries}
Currently, the vulnerability data were stored in raw CSV files and consolidated into a single JSON structure. While this approach was sufficient for the current stage of development, it was intended as a temporary solution. As additional datasets are incorporated, this structure is expected to become increasingly difficult to manage due to the growing size and complexity of the combined JSON file. At larger scales, searching, filtering, and updating records with a monolithic file would introduce performance limitations and reduce maintainability. For this reason, the transition toward a dedicated database system was identified as a necessary next step in the project's development. 

The implementation of a database system was intended not only to improve data organization and scalability, but also to enable more advanced interaction with the dataset through an AI-assisted query interface. This system would allow users to issue natural language or structured queries to explore relationships between vulnerability indicators and climate hazards without manually navigating the visualization layers. For example, users could query statements such as "Which census tracts with high elderly populations overlap with storm surge zones," "areas with high flood risk and low internet access," or "regions most affected by landslide risk and low household income."

By integrating an AI-driven query layer, the system would support more dynamic and intuitive exploration of spatial relationships, allowing users to identify patterns and correlations across large, multidimensional datasets more efficiently. This approach would reduce reliance on manual inspection of maps and tables, and instead provide a more direct analytical interface for decision-making and research applications.

\subsection{Scenario-Based Layer Filtering}
Another proposed enhancement to the system was the introduction of a scenario-based layer filtering mechanism. This feature was intended to improve interpretability by dynamically adjusting available layer options based on the selection of primary metric. Since certain vulnerability indicators and climate hazards exhibit stronger relationships than others, the system would prioritize relevant layers by highlighting or enabling strongly related datasets while visually de-emphasizing less relevant ones. 

When a user selected a specific metric or scenario, the interface would automatically filter the available layer options to reflect the most relevant correlations, supporting more guided and context-aware analysis. This approach was intended to reduce cognitive load and improve the clarity of relationships between datasets, particularly when working with many overlapping climate and socioeconomic variables. 

Importantly, this functionality was designed to remain optional. The existing sandbox-style configuration would be preserved, allowing users to retain full control over all layers when unrestricted exploration was preferred. In this way, the system would support both structured, scenario-driven analysis and open-ended exploratory use within the same framework. 

\subsection{Info blocks}
Another near-term enhancement involved improving the clarity and interpretability of individual datasets through the addition of contextual information features. While many of the existing metrics were self-explanatory based on their labels, several required additional clarification to ensure correct interpretation by users with varying levels of domain knowledge. Initial user and stakeholder feedback indicated that the lack of embedded explanations occasionally reduced usability and increased uncertainty when interpreting specific indicators. 

To address this, future iterations of the system were planned to incorporate informational tool tips or icon-based help features within the table and map interfaces. These elements would allow users to access concise descriptions of each metric through simple interactions such as hovering or clicking. The goal of this enhancement was to provide immediate contextual guidance without requiring users to leave the visualization environment or consult external documentation.

This improvement was intended to strengthen overall usability by ensuring that both technical and non-technical users could interpret datasets consistently and accurately, thereby supporting more informed analysis of both vulnerability and hazard-related information. 

\section{Conclusion}
This paper introduces the design and implementation of Community Census and Spatial Visualization Index (CCSVI). A unified geospatial platform that integrates climate hazard, socioeconomic, and infrastructure data for the state of Hawaiʻi. By combining these previously disjointed datasets within an interactive environment, the system allowed users to more effectively identify where environmental risks and social vulnerabilities intersect. 

The project demonstrated that through targeted data processing and optimization techniques, large and complex geospatial datasets can be transformed into responsive, web-based visualization system. The resulting platform supports real-time exploration, multi-map comparison, and layered analysis, allowing users to uncover spatial patterns that are difficult to detect through static methods. 

Ultimately, CCSVI highlights the value of accessible, integrated visualization tools in improving how climate risk is understood and communicated. By making these relationships more visible and interpretable, the system provides a practical foundation for more informed decision-making in disaster preparedness and climate adaptation.

\section{Acknowledgment}
This material is based upon work supported by the National Science Foundation Award No. 2417946, Strengthening the CyberInfrastructure Professionals Ecosystem (SCIPE) Cyberinfrastructure Pacific Professionals (CI-PP)
This work was enabled in part by funding from the National Science Foundation awards: 2149133, 2201428,  2232862, 2004014, 2003800, 2003387, 2117975, 2138259, 2138286, 2138307, 2137603, and 2138296.

\subsection{SCIPE Grant Students}
A special thank you to the undergraduates from the Strengthening the Cyberinfrastructure Professionals Ecosystem (SCIPE) grant who assisted with this project. Thank you to Steven Le\footnote{lesteven@hawaii.edu}, who helped with identifying dataset meanings and processed several datasets. Thank you to Jezza Michaella Villanueva\footnote{jmv9@hawaii.edu}, who initiated the creation of the database from the existing data.

\bibliographystyle{unsrt}
\bibliography{references}

\end{document}